\documentclass{INTERSPEECH2023}

\interspeechcameraready 

\usepackage{multirow}
\usepackage{cite}
\usepackage{float}
\usepackage{stfloats}
\title{DCTX-Conformer: Dynamic context carry-over for low latency unified streaming and non-streaming Conformer ASR}
\name{Goeric Huybrechts$^1$, Srikanth Ronanki$^1$, Xilai Li$^1$, Hadis Nosrati$^2$, Sravan Bodapati$^1$, Katrin Kirchhoff$^1$}
\address{
  $^1$AWS AI Labs, USA\\
  $^2$AWS AI Labs, Australia}
\email{huybrech@amazon.com, ronanks@amazon.com, lixilai@amazon.com, hnosr@amazon.com, sravanb@amazon.com, katrinki@amazon.com}

\begin{document}

\maketitle
 
\begin{abstract}

Conformer-based end-to-end models have become ubiquitous these days and are commonly used in both streaming and non-streaming automatic speech recognition (ASR). Techniques like dual-mode and dynamic chunk training helped unify streaming and non-streaming systems. However, there remains a performance gap between streaming with a full and limited past context. To address this issue, we propose the integration of a novel dynamic contextual carry-over mechanism in a state-of-the-art (SOTA) unified ASR system. Our proposed dynamic context Conformer (DCTX-Conformer) utilizes a non-overlapping contextual carry-over mechanism that takes into account both the left context of a chunk and one or more preceding context embeddings. We outperform the SOTA by a relative 25.0\% word error rate, with a negligible latency impact due to the additional context embeddings.


\end{abstract}

\section{Introduction}

Recently, end-to-end automatic speech recognition (ASR) systems such as attention-based encoder-decoder \cite{chan2016listen, chorowski2015attention}, CTC \cite{graves2006connectionist, amodei2016deep, dingliwal2023personalization} and Transducer \cite{graves2012sequence, graves2013speech} have become popular due to their simplicity in combining pronunciation, language and acoustic models into one neural network. Although these state-of-the-art (SOTA) models perform well in full-contextual (i.e. non-streaming) situations, they experience a decline in performance when used in real-time streaming scenarios due to the lack of future context \cite{dong2019self, zhang2020transformer, huang2020conv, miao2020transformer}. Likewise, performance degrades to an even greater extent when streaming with a limited instead of a full past context \cite{dong2019self, zhang2020transformer, huang2020conv, miao2020transformer}.

These two sources of streaming performance degradation hold both for purely streaming \cite{dong2019self, zhang2020transformer, huang2020conv, miao2020transformer, chiu2017monotonic, sainath2019two} and the more recent unified ASR systems \cite{tripathi2020transformer, hu2020deliberation, zhang2020unified, yu2021dual, yao2021wenet, moritz2021dual, dcconv}. The latter systems unify streaming and non-streaming into a single model, which helps reduce development, training and deployment cost. While some studies like \cite{an2022cuside} have explored solutions to mitigate the lack of future context, we focus on the challenge of limited past context as we are particularly interested in low latency systems. Similarly, we constrain ourselves to unified ASR systems for their advantages highlighted earlier.

A commonly explored solution to overcome the limitation of a restricted past context in block-processing models \cite{dong2019self} is to store and propagate the history context. Tsunoo et al. \cite{tsunoo2019transformer} introduce a context-aware inheritance mechanism in the self-attention layers of a Transformer \cite{vaswani2017attention}. A context embedding is appended as extra frame to each chunk before the self-attention and is handed over from one chunk/layer to another to help encode not only local acoustic information but also global linguistic, channel and speaker attributes. Related works consider multiple history embeddings with the introduction of memory banks. The self-attention unit of the Augmented Memory Transformer \cite{wu2020streaming} attends on a short segment of the input sequence and a bank of memories that stores the embedding information from all previous processed segments. In Emformer \cite{shi2021emformer}, the long-range history context is distilled into an augmented memory bank in between the self-attention and feed-forward layers. While all these works made some great progress, there still exists a gap between streaming with a full and limited past context. Plus, none of these works have considered the unified streaming and non-streaming setting.

In this work, we tackle the streaming performance degradation for unified ASR systems when using a limited chunk's left context. We propose the dynamic context Conformer (DCTX-Conformer) that builds on and enhances the SOTA with next 5 contributions: (1) We incorporate the contextual carry-over (CCO) mechanism of \cite{tsunoo2019transformer} in a SOTA unified ASR Conformer system \cite{dcconv}; (2) We improve upon the CCO mechanism by integrating a dynamic dependency on a chunk's left context; (3) We improve upon the CCO mechanism by adding a dynamic dependency on the number of context embeddings; (4) We conduct experiments using the dynamic CCO mechanism in the lower latency, non-overlapping streaming mode without any look-ahead frames; (5) We conduct an exhaustive experimental study of our model on different chunk sizes, various chunk's left contexts and multiple context embeddings. The results on numerous datasets and many different settings demonstrate the effectiveness and robustness of our proposed model.

\section{Approach and related work}

In this work, we improve upon the CCO mechanism of \cite{tsunoo2019transformer} and integrate it in a SOTA unified ASR Conformer system \cite{dcconv}. We consider a joint CTC\cite{graves2006connectionist}-attention framework \cite{kim2017joint, watanabe2017hybrid} for training our unified models. 

\subsection{End-to-end unified ASR}

For unified ASR models to perform well in both streaming and non-streaming settings, they must be exposed to both limited (i.e. streaming) and full (i.e. non-streaming) contexts during training. To accomplish this, \cite{yao2021wenet, dcconv} propose a dynamic chunk training (DCT) for self-attention layers which involves varying the chunk size dynamically at training time. As in \cite{dcconv}, we randomly sample a chunk size between 8 (= 320ms) and 32 (= 1280ms) down-sampled self-attention frames 60\% of the time and run a full-contextual training the remaining 40\%. Moreover, we use the same dynamic left context mechanism that allows to vary the left context between zero and all preceding chunks so that the model becomes robust to numerous left context sizes at inference time. The downside of \cite{dcconv} and other (unified) streamable systems is that there remains a non-negligible gap between streaming with a full and limited past context. To overcome this limitation, we integrate our proposed dynamic CCO mechanism that we discuss in next subsection.

\subsection{Dynamic contextual carry-over mechanism}

\begin{figure}[t]
  \centering
  \includegraphics[width=0.78\linewidth]{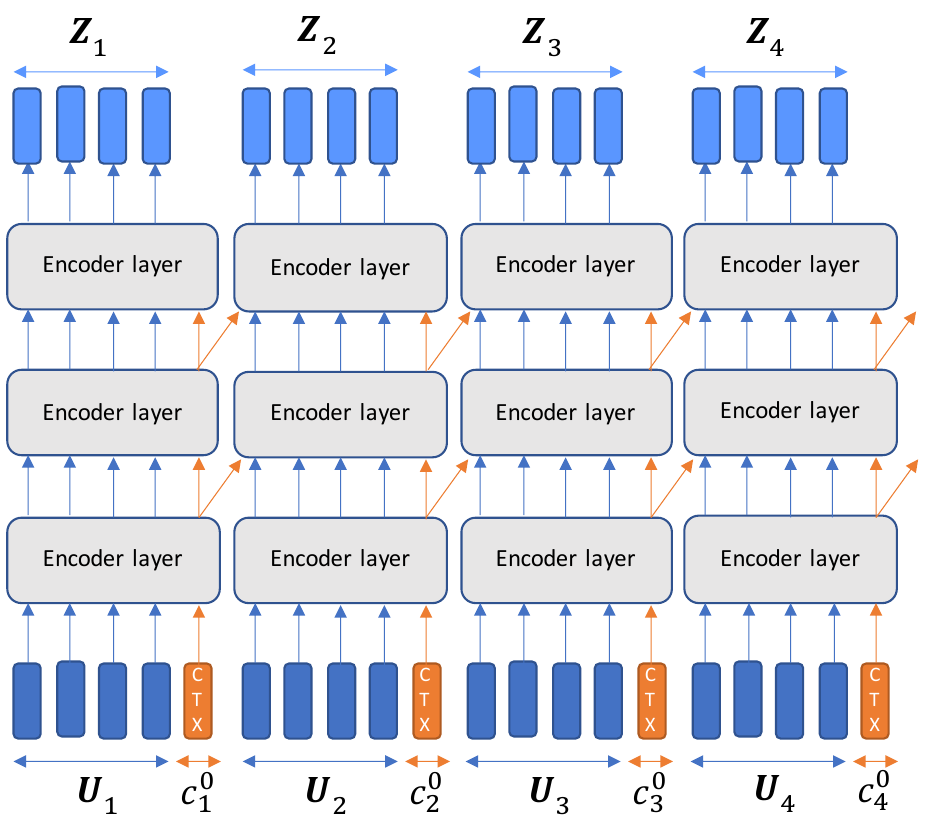}
  \caption{Contextual carry-over mechanism using non-overlapping chunks.}
  \label{fig:architecture}
\end{figure}

The authors of \cite{tsunoo2019transformer} extend the streaming Transformer model with a context-aware inheritance mechanism (Fig. \ref{fig:architecture}). Context embeddings are passed on from one layer/chunk to the next. An experimental study in \cite{tsunoo2019transformer} shows that taking the average of each chunk as context embedding in the first layer provides the best results. As opposed to the original work \cite{tsunoo2019transformer}, we integrate the CCO mechanism in a unified Conformer trained with dynamic chunk sizes. The Conformer is generally considered to be a better alternative than the Transformer for (unified) streaming ASR \cite{gulati2020conformer}, while the DCT approach makes the model more robust to different chunk sizes at inference time. Lastly, unlike \cite{dcconv, tsunoo2019transformer} that perform block processing with overlapping chunks, we stream in the lower latency non-overlapping manner. Besides these differences in architecture and applications, we make 2 contributions to the mechanism.

\begin{figure}[t]
  \centering
  \includegraphics[width=0.8\linewidth]{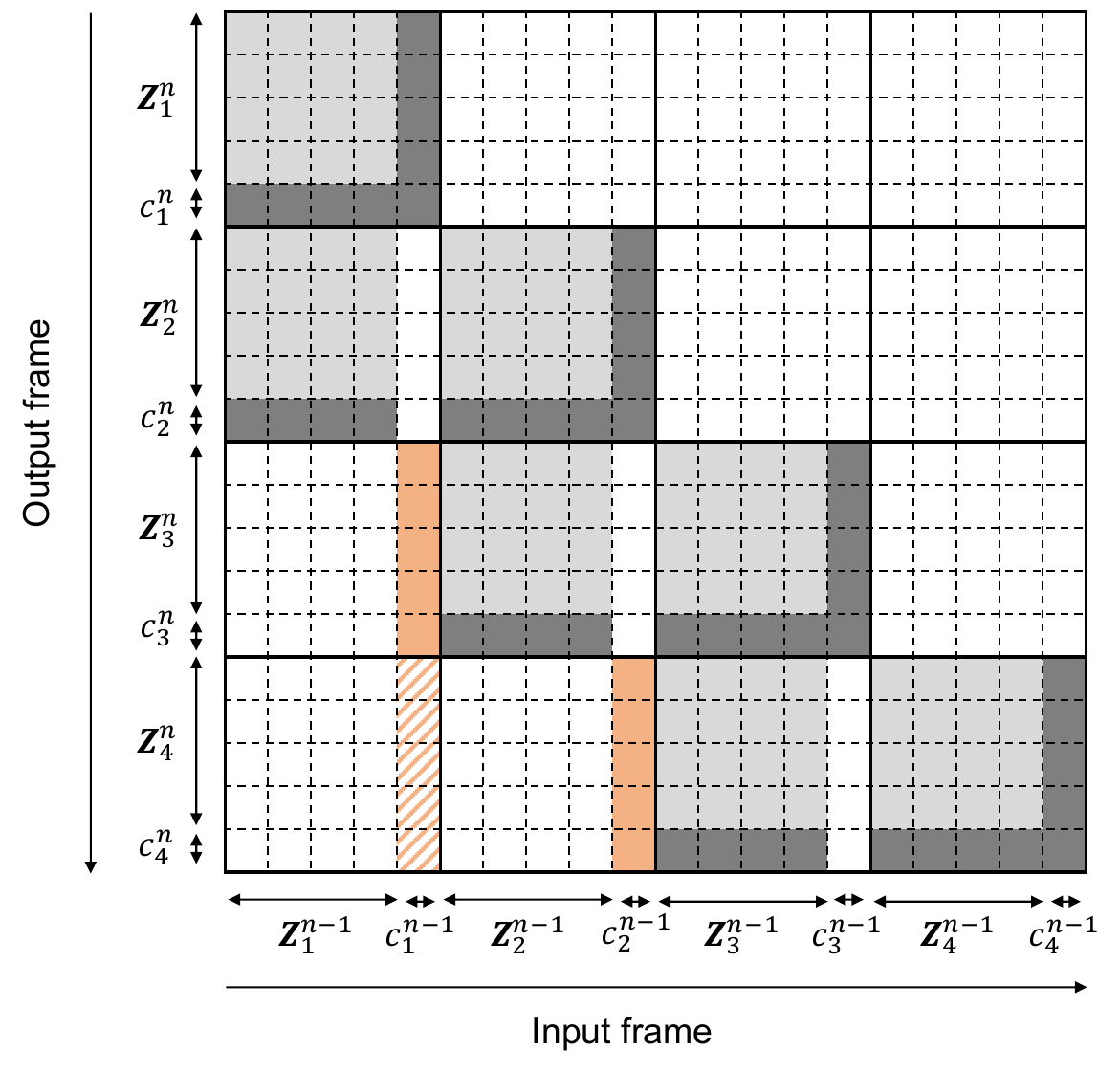}
  \caption{Contextual carry-over mask. Illustration for 4 non-overlapping chunks of size 4 with left context size set to 1 chunk. Non-white squares = 1, white squares = 0. Light gray = frame to frame dependency, dark gray = context embedding involved. Orange squares represent the context carried over from past context embeddings in the encoder layers $n > 1$.}
  \label{fig:masks}
\end{figure}

Firstly, we propose to keep a dynamic dependency on preceding chunks despite the presence of context embeddings. We demonstrate that this combination leads to significant performance gains for insignificant latency drops. The dynamic dependency consists in varying the chunk's left context size at training time. The context embedding is in that case handed over from the chunk that precedes the left context chunks and no longer from the one directly preceding the current chunk. This dynamic training trick allows to vary the amount of left context needed at inference time depending on the application's latency requirements. Fig. \ref{fig:masks} shows the design of the self-attention mask in the use-case of 4 non-overlapping chunks of size 4 with the left context size set to 1 chunk. We formalise the novel dynamic CCO process next. Let $\textbf{U}_{i}$ denote the down-sampled chunks that are passed to the first self-attention layer, and $c_{i}$ denote the corresponding context embeddings. The attention computation takes the embedding dimension \textit{d}, queries \textbf{Q}, keys \textbf{K}, values \textbf{V} and \textit{Mask} of Fig. \ref{fig:masks} as input:

\vspace{-5mm}
\begin{equation}
Attention(\textbf{Q}, \textbf{K}, \textbf{V}) = Softmax(\frac{Mask(\textbf{Q}\textbf{K}^T)}{\sqrt{d}})\textbf{V},
\end{equation}
with the \textit{Mask} adapting \textbf{Q}, \textbf{K} and \textbf{V} as follows:

For layer 1:
\begin{equation}
\textbf{Q}^1_b = [\textbf{U}_b, c^0_b],
\end{equation}
\begin{equation}
\textbf{K}^1_b = \textbf{V}^1_b = [\textbf{U}_{b-LC:b}, c^0_b],
\end{equation}
where $b$ denotes the chunk (or block) number, $LC$ denotes the number of left context chunks and $\textbf{U}_{b-LC:b}$ denotes the concatenation of chunks $\textbf{U}_{b-LC}$ to $\textbf{U}_b$ (included).

For layer n $>$ 1:
\begin{equation}
\textbf{Q}^n_b = [\textbf{Z}^{n-1}_b, c^{n-1}_b],
\end{equation}
\begin{equation}
\textbf{K}^n_b = \textbf{V}^n_b = [c^{n-1}_{b-LC-1}, \textbf{Z}^{n-1}_{b-LC:b}, c^{n-1}_b],
\end{equation}
where $\textbf{Z}^n_b$ is the output of the $n^{th}$ encoder layer of chunk (or block) $b$.

Secondly, inspired by the idea of memory banks \cite{wu2020streaming, shi2021emformer}, we show further improvements by relying on more than one preceding context embedding at inference time. For instance, in Fig. \ref{fig:masks}, output chunk \#4 would depend on all frames of chunks \#3 and \#4 and context embeddings \#1 (dashed orange squares), \#2 (orange squares) and \#4 (dark gray squares). This additional dependency on the preceding context embedding \#1 is set at inference time only, as opposed to \cite{wu2020streaming, shi2021emformer}. The self-attention for chunk $b>LC+1$ and layer $n>1$ is still trained relying on one preceding context embedding $c^{n-1}_{b-LC-1}$ only (i.e. the dashed orange squares are set to 0 at training time). If we assume to depend on $N_{ctx}$ preceding (successive) context embeddings, the calculation of the keys and values becomes:
\begin{equation}
\textbf{K}^n_b = \textbf{V}^n_b = [c^{n-1}_{b-LC-N_{ctx}:b-LC-1}, \textbf{Z}^{n-1}_{b-LC:b}, c^{n-1}_b]
\end{equation}

Unlike \cite{wu2020streaming, shi2021emformer}, our queries only depend on the actual chunk and not on any patched left or right context. Plus, our keys and values do not contain any redundant history information as we only include the context embeddings that summarize the context before the chunk's patched left context. Similarly to the keys, they do not contain any patched right context. More importantly, in \cite{wu2020streaming, shi2021emformer} memory bank entries are recomputed at every layer as an average projection of the chunk. In our CCO mechanism, context embeddings are only initialised as a chunk's average in the first layer. This gives the model more freedom to learn superior contextual representations in subsequent layers. Every intermediate contextual embedding $c^{n}_b$ also explicitly depends on context embeddings $c^{n-1}_{b-LC-1}$ and $c^{n-1}_b$, allowing to better model interactions between them than the memory banks in \cite{wu2020streaming, shi2021emformer} where this explicit interaction does not exist.

\section{Experimental settings}

\subsection{Datasets}

\noindent\textbf{Training} We consider 3 different speech corpora varying in size for training our models: (1) The open-source \textit{LibriSpeech} \cite{panayotov2015librispeech} corpus, for which we combine \textit{train-clean-100}, \textit{train-clean-360} and \textit{train-other-500} to have 960 hours of training data; (2) A \textit{small-scale} 1k hour English corpus and (3) a \textit{large-scale} 10k hour superset, sampled from in-house paired audio and text data. Both corpora include audio files with a good mix of accents, speakers, sampling rates and background noise. These 3 data regimes are representative of a wide range of end-to-end ASR systems for various speech applications.

\noindent\textbf{Evaluation} For the \textit{LibriSpeech} experiments, we evaluate our models on \textit{test-clean} and \textit{test-other}, whose average utterance length is 7s. For the \textit{small-} and \textit{large-scale} experiments we use the following diverse public test sets: (1) \textit{MTDialogue\footnote{https://github.com/Phylliida/Dialogue-Datasets}}: Collection of movie and Twitter data. The dataset is 1.2h long and the average utterance length is 3s; (2) \textit{Wall Street Journal (WSJ)}: We use WSJ's eval\_test92 \cite{garofolo1993csr}, prepared using Kaldi's \cite{povey2011kaldi} WSJ recipe. The dataset is 0.7h long and the average utterance length is 8s; (3) \textit{Voxpopuli} \cite{wang2021voxpopuli}: We use the English test partition. The dataset is 4.9h long and the average utterance length is 10s. 

\subsection{Setup}

\noindent\textbf{Training} We use a Conformer as the encoder and a shallow single-layer Transformer as the attention-based decoder. The inputs are 80 dimensional log-mel features extracted with 25ms FFT windows and 10ms frame shifts. For the \textit{LibriSpeech} experiments, we use a Conformer-12x512x8, which consists of 12 encoder layers with 512 feature dimensions and 8 self-attention heads. We use ESPnet's pre-trained 24-layered Transformer-based neural LM on the \textit{LibriSpeech-train} dataset for rescoring. For the \textit{small-scale} experiments, we use a Conformer-16x256x4 and a BPE embedding of size 1024. For the \textit{large-scale} experiments, we use a Conformer-16x512x8 and a BPE embedding of size 2048. We train a 4-gram LM on the training text for shallow fusion. The kernel size of our convolution modules is 31. We optimise our model via the hybrid CTC and attention losses. All of our models are trained for 60 epochs with the Adam optimizer \cite{kingma2014adam} and a warm-up learning rate scheduler. The unified ASR models are fine-tuned with DCT for 30 epochs from a full-contextual model trained for 30 epochs, as is done in \cite{dcconv}. We make use of ESPnet \cite{watanabe2018espnet} and p4de.24xlarge Amazon EC2 instances that consist of 8 NVIDIA A100 Tensor Core GPUs.

\noindent\textbf{Evaluation} We discard the attention-based decoder and use the CTC decoder to generate outputs with a CTC prefix beam search and beam size of 20 for the \textit{LibriSpeech} experiments and of 50 for the \textit{small-} and \textit{large-scale} experiments. A CTC decoder optimises the real-time factor compared to the attention-based decoder, as the latter is non-autoregressive and needs triggered attention \cite{moritz2019triggered} for streaming inference. All of our streaming results are obtained via non-overlapping streaming without any look-ahead context, unless otherwise stated.

\begin{table*}[t]
\caption{WER for full-contextual setting and for non-overlapping streaming without look-ahead frames in function of chunk size (ms).}
\vspace{-2mm}
\label{tab:chunksize}
\centering
\footnotesize
\resizebox{1.0\textwidth}{!}{
\begin{tabular}{l|l|llll|l|llll|llll|llll}
\toprule
\multirow{2}{*}{Model}                             & \multirow{2}{*}{Test set} & \multicolumn{4}{c|}{\textbf{LibriSpeech}} & \multirow{2}{*}{Train set} & \multicolumn{4}{c|}{\textbf{MTDialogue}} & \multicolumn{4}{c|}{\textbf{WSJ}} & \multicolumn{4}{c}{\textbf{VoxPopuli}} \\
                                                   &                                                                                 & Full   & 1280   & 640   & 320   &                            & Full   & 1280   & 640   & 320  & Full & 1280 & 640 & 320 & Full   & 1280   & 640  & 320  \\
\midrule
(A) Full-contextual, LC=0                                 & \multirow{5}{*}{Test-clean}                                                     &  2.1      &  22.5      &  49.4     &  89.0     & \multirow{5}{*}{Small}                      &  9.5      & 14.8       & 23.9      & 46.8     & 7.8     & 18.3     & 31.9    & 57.9    & 14.5       & 26.4       & 42.8     & 68.8     \\
(B.1) Conformer w/o CCO, LC=Full                       &                                                                                 &  2.1      & 2.4       &  2.6     &  3.1     &                            &        9.3   & 9.8      & 10.4     & 16.6     & 8.2   & 9.0      & 11.3     & 35.4     & 14.7   & 16.8      & 21.4     & 56.0      \\
(B.2) Conformer w/o CCO, LC=0                          &                                                                                 &   2.1     &   2.8     &  3.6     &   5.4    &                            &        9.3  & 10.3      & 14.2     & 48.7     & 8.2   & 10.7      & 18.3     & 80.5     & 14.7   & 19.5      & 34.6     & 85.9      \\
(C) \textbf{DCTX-Conformer, LC=0} &                                                                                 &  2.3      &  2.6      &  3.2     & 4.4      &                            &        9.4   & 10.7      & 13.1     & 25.1     & 8.3   & 10.5      & 14.4     & 38.6     & 14.8   & 19.0      & 25.3     & 63.5      \\
\midrule
(A) Full-contextual, LC=0                                 & \multirow{5}{*}{Test-other}                                                     &  5.1      &  29.5      &  58.2     & 92.7      & \multirow{5}{*}{Large}                      & 7.0       & 12.2       &  20.0     & 39.8     &  5.2    &  13.2    & 24.5    & 50.9    & 10.2       & 19.0       & 33.9     &  60.8    \\
(B.1) Conformer w/o CCO, LC=Full                       &                                                                                 &   4.9     &  6.0      &  6.8     & 8.2      &                            &        6.8   & 7.2      & 8.7     & 22.0     & 5.1   & 5.7      & 10.3     & 43.9     & 10.1   & 11.6      & 17.4     & 57.2      \\
(B.2) Conformer w/o CCO, LC=0                          &                                                                                 &  4.9      & 7.4       &  9.6     & 14.8      &                            &        6.8   & 8.4      & 12.5     & 33.2     & 5.1   & 8.4      & 15.5     & 58.0     & 10.1   & 14.9      & 21.7     & 75.0      \\
(C) \textbf{DCTX-Conformer, LC=0} &                                                                                 &  6.0      &  6.8      & 8.8      &  12.8     &                            &        6.7   & 7.9      & 9.6     & 16.8     & 5.0   & 6.5      & 12.2     & 27.2     & 10.1   & 13.6      & 20.3     & 40.5   \\
\bottomrule
\end{tabular}}
\end{table*}

\section{Results}

\subsection{Performance impact of chunk size}

In Table \ref{tab:chunksize}, we analyze the impact of the chunk size on the word error rate (WER) by comparing 3 types of models on 5 test sets: (A) A purely non-streamable model trained in a full-contextual setting (i.e. no DCT); (B) A unified SOTA model \cite{dcconv} without CCO and with full (B.1) and no (B.2) left context at inference time; (C) Our unified model with dynamic CCO and no left context at inference time. All models share the same architecture and were trained on the \textit{small-scale}, \textit{large-scale} and \textit{LibriSpeech} datasets. The results show that our model (C) roughly maintains the full-contextual performance of models (A) and (B), except for the LibriSpeech \textit{test-other} dataset, while significantly reducing the streaming performance gap between (B.1) and (B.2). We observe that the relative improvements increase as the chunk size decreases because context embeddings are increasingly impactful for smaller and therefore inferior acoustic representations. For the \textit{large-scale} model with inference chunk size 320ms, we even exceed the gap. Our hypothesis is that context embeddings provide a more effective way of incorporating past context than utilizing all frames from a full past context input. Moreover, we believe that the larger training set leads to a better modelling of those extra embeddings. Their use also significantly reduces the computational memory over non-contextual models that stream with full past context as the number of keys and values in the self-attention calculations decreases substantially. Overall, we notice an average gap reduction of 24.3\%, 49.5\% and 109.2\% across all datasets and models between no and full past context when streaming with a chunk size of 1280ms, 640ms and 320ms respectively.

\subsection{Performance impact of left context size}

\begin{figure}[th]
    \centering
    \includegraphics[width=1.0\linewidth]{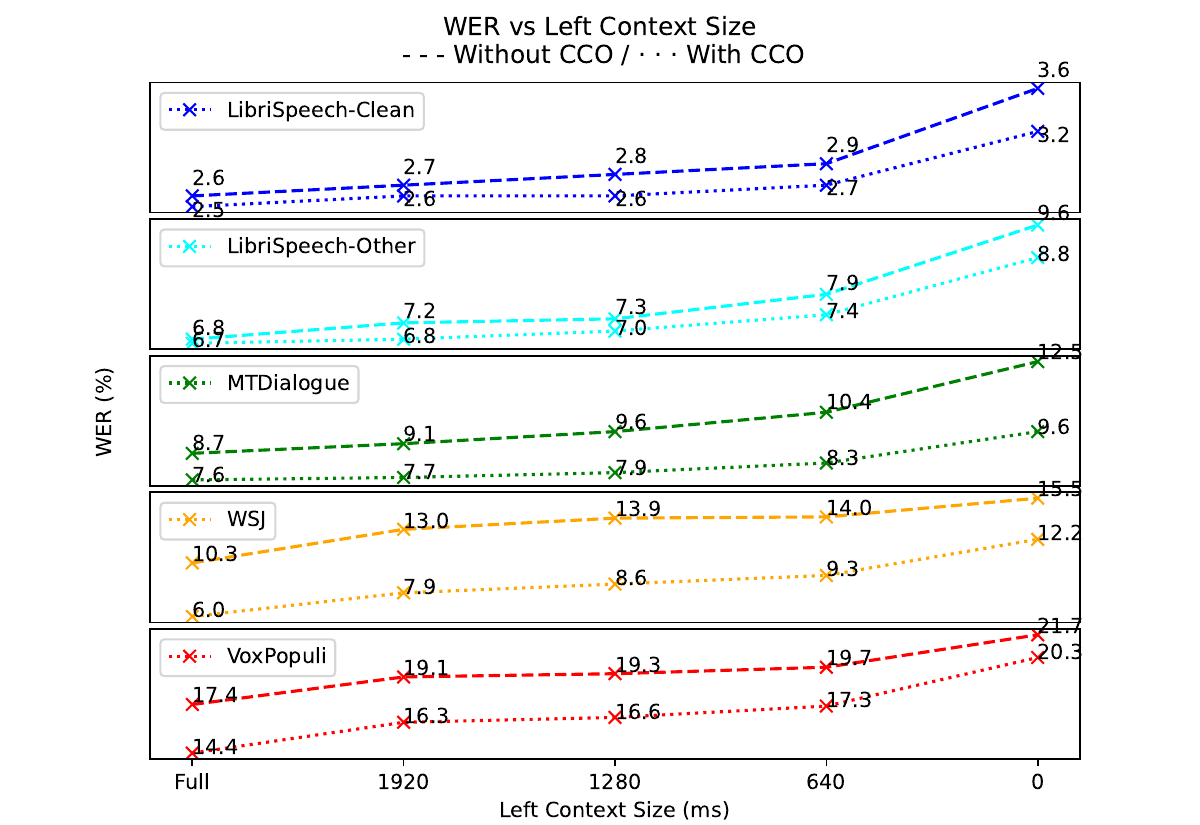}
    \caption{WER in function of left context (ms) for a chunk size of 640ms without look-ahead frames for model without and with context carry-over.}
    \label{fig:leftcontext}
\end{figure}

As can be observed in Table \ref{tab:chunksize}, the streaming gap between the model without CCO and with full left context (B.1) and our model with dynamic CCO and with no left context (C) can still be further reduced in most settings. In Fig. \ref{fig:leftcontext}, we analyze the impact of adding a chunk's left context on top of the CCO mechanism for the \textit{LibriSpeech} and \textit{large-scale} models when running inference with a chunk size of 640ms. The graphs indicate that the aforementioned gap not only further decreases as we add more left context, but also that with only 1 left context chunk we now outperform the \textit{large-scale} model (B.1) with full left context (as we already did in Table \ref{tab:chunksize} with our \textit{large-scale}  model for a chunk size of 320ms and left context size 0). Taking a closer look at \textit{WSJ} in particular, we observe that taking 3 left context chunks instead of none while still carrying over context leads to a WERR of 35.2\% and 49.0\% over the model with and without context carry-over respectively. As our models are trained with a dynamic left context, it gives the user the flexibility to easily adjust the left context size at inference time depending on the latency and memory requirements.

\subsection{Performance impact of context embeddings}

\begin{figure}[th]
    \centering
    \includegraphics[width=1.0\linewidth]{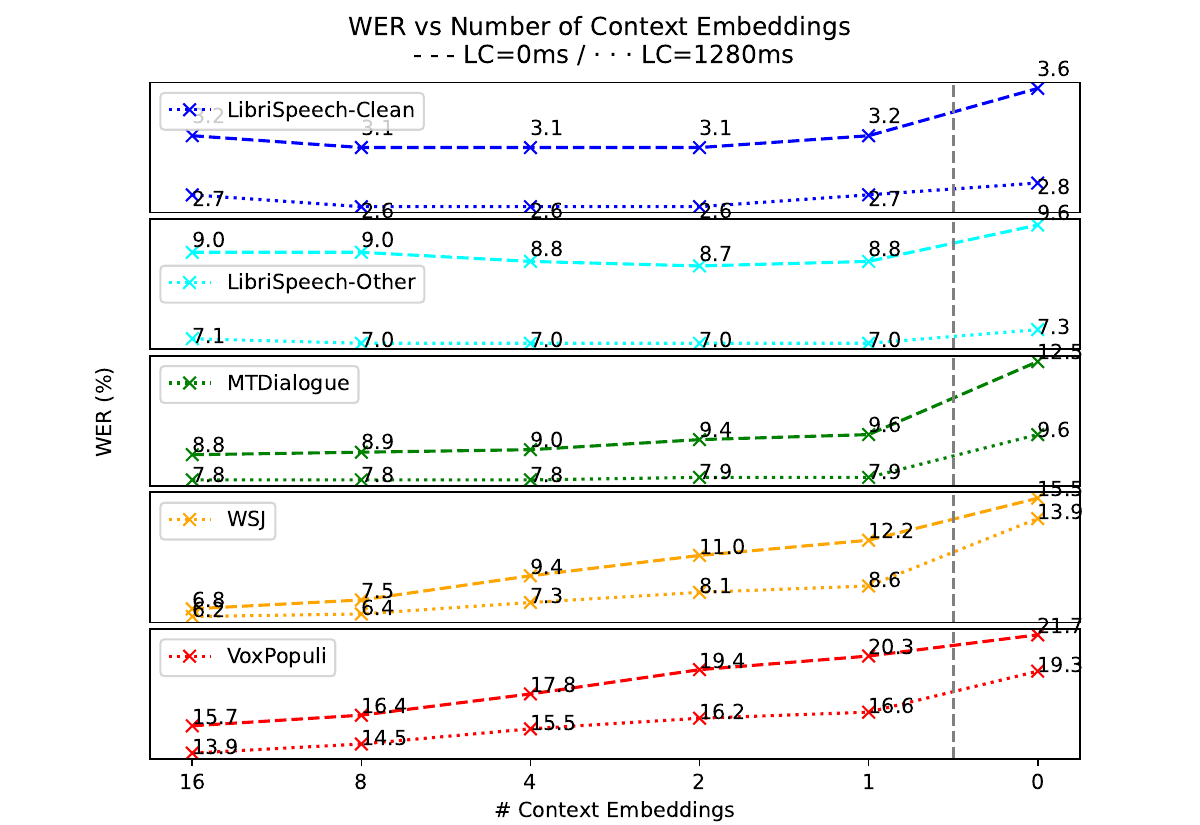}
    \caption{WER in function of number of context embeddings for a chunk size of 640ms without look-ahead frames and with left context of 0ms and 1280ms. For 0 context embeddings, we consider the baseline model without context carry-over.}
    \label{fig:nctxembeddings}
\end{figure}

In Fig. \ref{fig:nctxembeddings}, we showcase that relying on more than 1 preceding context embedding substantially improves the WER. We observe those improvements despite the fact that the model was trained with 1 preceding context embedding only. The more context embeddings we provide each chunk, the better the model performs, except for the \textit{LibriSpeech} test sets where a minor degradation of 0.9\% is observed across the two depicted left context settings when using 16 context embeddings instead of 1. For \textit{WSJ} and \textit{VoxPopuli} on the other hand, we observe an average WERR improvement of 36.1\% and 19.5\% respectively. Compared to the baseline Conformer model without CCO those numbers even increase to 55.8\% and 27.8\%. For \textit{MTDialogue}, we notice a faster saturation and only notice slight improvements beyond 4 context embeddings because the average utterance length in that dataset is small (= 3s). Therefore, for most utterances the model can only rely on a few past context embeddings. Overall, we demonstrate that our approach leads to an average 25.0\% WERR improvement over the non-contextual baseline model  across the datasets and the two depicted left context settings when using 16 context embeddings.

\subsection{LibriSpeech comparison with SOTA}

\begin{table}[t]
\caption{LibriSpeech WER comparison with SOTA look-ahead context overlapping streaming models.}
\vspace{-3mm}
\label{tab:librispeech}
\centering
\footnotesize
\resizebox{0.47\textwidth}{!}{
\begin{tabular}{l|ccccc|cc}
\toprule
Model                               & \multicolumn{1}{c}{LC} & \multicolumn{1}{c}{CC} & \multicolumn{1}{c}{RC} & \multicolumn{1}{l}{$N_{ctx}$} & \multicolumn{1}{c|}{\# Params} & \multicolumn{1}{c}{Clean} & \multicolumn{1}{c}{Other} \\
\midrule
AM-Transformer \cite{wu2020streaming} & 640                            & 1280                         & 320                             & Inf                       & 80M                                  & 2.8                            & 6.7                            \\
Emformer \cite{shi2021emformer}                     & 1280                           & 640                          & 320                             & 4                         & 120M                                 & 2.6                            & \textbf{6.0}                            \\
Conformer w/o CCO \cite{dcconv}          & 1280                           & \textbf{320}                          & 320                             & 0                         & 115M                                 & 2.5                            & 6.6                            \\
\textbf{DCTX-Conformer, w/o RC}         & 1280                           & 640                          & \textbf{0}                               & 2                         & 115M                                 & 2.6                            & 7.0     \\
\textbf{DCTX-Conformer, w/ RC}         & 1280                           & \textbf{320}                          & 320                               & 2                         & 115M                                 & \textbf{2.4}                            & 6.4 \\
\bottomrule
\end{tabular}
}
\vspace{-5mm}
\end{table}

In Table \ref{tab:librispeech}, we compare our model (with 2 context embeddings) to the Augmented Memory Transformer \cite{wu2020streaming} (with infinite memory bank entries and \textit{WAS} \cite{shi2020weak}), the Emformer hybrid system \cite{shi2021emformer} (with 4 memory bank entries and \textit{SMBR} \cite{vesely2013sequence}) and the baseline Conformer model without CCO \cite{dcconv} on the \textit{LibriSpeech} test sets. We provide numbers given in the respective papers in similar settings, where numbers were however given for the higher latency overlapping streaming mode with look-ahead/right context (RC). As demonstrated in \cite{zhang2020transformer}, while no look-ahead context improves latency, it significantly degrades the WER on the \textit{LibriSpeech} test sets. Despite this, our overall smaller segment (=LC+CC+RC) and our model being trained in a unified fashion, our lower latency model performs equally well on \textit{test-clean} and is competitive on \textit{test-other} wrt the SOTA. When we do use look-ahead context, we even outperform every SOTA model on \textit{test-clean} and are just behind Emformer on \textit{test-other}, even though our model did not see any look-ahead context during training and uses only half of the Emformer center context (CC) size.

\vspace{-2mm}
\subsection{Latency study}

In Table \ref{tab:latency}, we provide some latency results in function of the number of context embeddings using the \textit{small-scale} models, a 640ms chunk, a 1280ms left context and the \textit{VoxPopuli} dataset. The measurements indicate the negligible latency impact of the context embeddings.

\vspace{1mm}
\begin{table}[h]
\caption{Latency in ms in function of number of context embeddings for model without and with context carry-over.}
\vspace{-3mm}
\label{tab:latency}
\centering
\footnotesize
\resizebox{0.45\textwidth}{!}{
\begin{tabular}{lcccc}
\toprule
                      & \textbf{Conformer w/o CCO} & \multicolumn{3}{c}{\textbf{DCTX-Conformer}} \\
\cmidrule{2-5}
\# Ctx embeddings & 0                 & 1          & 8         & 16        \\
\midrule
Mean    & 816               & 820        & 820       & 821       \\
P99       & 954               & 960        & 960       & 961  \\
\bottomrule
\end{tabular}}
\end{table}

\vspace{-6mm}
\section{Conclusions}

In this work, we incorporate an improved version of the contextual carry-over mechanism in a state-of-the-art unified ASR system. We modify the contextual carry-over mechanism by integrating a dynamic dependency on both the chunk's left context size and preceding context embeddings. With an exhaustive experimental study on many datasets, we show the efficacy and robustness of our proposed approach. The results demonstrate that our DCTX-Conformer model more effectively captures a full past context with reduced latency and computational memory usage in streaming scenarios, without compromising its non-streaming performance.

\bibliographystyle{IEEEtran}
\bibliography{mybib}

\end{document}